\documentclass[aps,pre,twocolumn,floatfix,showpacs,showkeys]{revtex4}
\usepackage{amsmath,amsfonts,amssymb,mathrsfs,graphicx}
\begin{document}
\title{Neural Network with Semiclassical Excitation}

\author{Fariel Shafee}
\affiliation{ Department of Physics\\ Princeton University\\
Princeton, NJ 08540\\ USA.} \email{fshafee@princeton.edu }

\begin {abstract}

We have constructed a simple semiclassical model of a neural network
where the neurons have quantum links and affect one another in a
fashion analogous to action potentials. However, there are
significant differences with the classical models in the nonlinear
functional dependence of the probability of firing of a neuron on
the superposed signal from the neighbors and also the
nondeterministic fundamental stochasticity of the quantum process.
Remarkably average periodicity is nevertheless observed in
simulation, which agrees quite well with a formula derived from the
nonlinear relationship. Short-term retentivity of input memory is
observed to be subtle but perceptible. This suggests the use of such
a network where short-term dynamic memory may be a desirable
objective.
\end{abstract}

\vspace*{1cm}

\maketitle

\section{Introduction}

The classical integrate-and-fire neural network has been studied
both in the simpler zero-width \cite{HH1} action potential and the
more involved finite-width \cite{FS1}cases. In these works every
neuron integrates the current coming from neighboring neurons and
when the threshold for firing is exceeded, it too sends off an
action potential to its neighbors. Hopfield and Herz had found that
there is a simple relation between the contributions  $A$ from the
neighbors and an external current $I$, with the time period of the
firing of the network when phase-lock is established:

\begin{equation} \label{eq1}
\tau = (1-A)/I
\end{equation}

It has been shown \cite{FS1} that there is some modification of this
formula when the action potential is not exactly a delta function
but is spread over time, which is, of course, a more realistic
assumption, in biological as well as physical contexts.

In view of the recent great interest in quantum computing we think
it is worthwhile to investigate the changes if any that would result
from converting such networks to a quantum model. As a first step,
we have here tried to see the changes when the action potential acts
like a quantum radiation to the neighbors, instead of a current, and
the firing of the neuron is replaced by induced radiation from it to
the neighbors. The quantity integrated therefore gives not a
classical state with deterministic firing on reaching the threshold,
but the superposed quantum amplitude, the square of whose magnitude
is related to the probability of firing.

Hence, instead of the deterministic approach of the classical case,
we have a quantum stochastic interpretation of the dynamics of the
network. It is obvious that if the neurons are allowed to emit
according to time dependent perturbation theory at random, there
will be no exact phase-locking, but it may still be possible to find
average periods of firing of a typical neuron which may depend on
the strengths of the various parameters of the model, as in the
classical case. We study this important possibility in this paper,
because it may throw light on some of the most fundamental
similarities and differences between classical and quantum networks.

This model is not a fully quantized network where the states of all
the neurons are entangled and the operations are all unitary, as in
quantum computing models. Our interest is not in the exploitation of
the enlargement of the capacity of the net in the quantum form, or
of the speed of information processing in such a system, but to make
an investigation of the significant changes that the nonlinearity
and the stochasticity may be expected to bring about. The problems
of measurement make the attractive possibilities of a fully
quantized networked system of gates and other devices  (see e.g.
\cite{CN1}) unreachable in the near future. Hence, as a first step
it is possible that quantum computers will make a beginning as
semiclassical devices with all links not fully entangled by a
unitary operation. How dissipation destroys the memory of the
initial state and how quickly, and what periodicity, if any,
emerges, depending on which parameters, are important questions that
need to be studied.

\section{The Model}

We have simulated a square lattice of such neurons which can
communicate with neighbors with quantum signals and have varied
their durations, in analogy with the action potential currents of
varying widths in the classical case. We have also retained a
constant potential background in analogy with the constant
external current.

We know
\begin{equation} \label{eq2}
|t> = U(t,t_0)|t_0>
\end{equation}

with

\begin{equation} \label{eq3}
U(t,t_0) = exp[i \int{dt V}] \approx 1+ i  \int{dt V}
\end{equation}

So that the transition rate

\begin{equation} \label{eq4}
\Gamma   \approx  |  \int{dt V }|^2
\end{equation}

In our case we have retained a constant $V_o$ background and have
added the contributions  $ \int{dt f(t_i) v }$   from the $i-th$
neighbor. The integration is spread over the duration of the
quantum pulse since the last triggering.

We might expect in this case

\begin{equation} \label{eq5}
\tau= 1/ |\int{ dt (V_o + \Sigma{ f_i v})} |^2
\end{equation}

For small perturbations this would look like

\begin{equation} \label{eq6}
\tau = k'( 1- |\Sigma{\int{dt f_i v} |^2)}/|\int {dt V_o}|^2
\end{equation}

which is analogous to the classical formula \cite{FS1,HH1}.

However, we must realize that in the classical case it is possible
to have    $1 = A$ (in the standardized units used there) and get
a zero period, i.e. an always saturated net. This is impossible in
this quantum case because the more exact expression (1) cannot
give a zero  .

If we make the further assumption that the pulse width is a duration
greater than the average period, then we have to take a
proportionate amount of contribution from the neighbors. In this
case we get the relation, for a square lattice network:

\begin{equation} \label{eq7}
1/\tau = k' ( V_0 \tau+ 4 v \tau/w)^2
\end{equation}

which is a cubic equation in the period and can be solved in terms
of the system parameters $V_0$ , width $w$ of the pulse and the
pulse size $v$. The nonlinearity of the quantum version of the
network mentioned earlier is responsible for this form.

Unlike the classical case we have a randomness in the contribution
from the neighbors. We may try to account for this by introducing a
parameter $q$ with the interaction $v$ between neighbors.  This
gives:

\begin{equation} \label{eq8}
1/\tau = k ( V_0  + 4 q v /w)^{(2/3)}
\end{equation}

\section{Input Dependence}

A system that does not respond to the environment is useless. So
while the system period found above has its intrinsic interest, we
need also study the behavior of the system to different inputs,
i.e. initial states of pre-assigned nodes to serve as the
interface with the environment. Obviously the system behavior will
depend on the geometry and connectivity of the net and the pattern
of the input. However, in this semi-quantum model we foresee a
loss of memory as the nodes fire and collapse from their original
states. This may be a desirable characteristic if we are
interested in a system with only a short-term memory, because the
automatic erasure of memory after a time may save us the software
steps or the hardware needed to erase.

 We shall principally study simulation experiments, as the details
 of memory loss are expected to depend in a complicated way on the system
 used and any simple closed analytic expression is unlikely.
 However, we shall also try to see of a simple formula such as Eq.
  \ref{eq8} can give agreements with the simulation for the
  average periods of oscillation after the initial memory is
  effaced.

\section{Results of Simulation}

\begin{figure}[ht!]
\includegraphics[width=8cm]{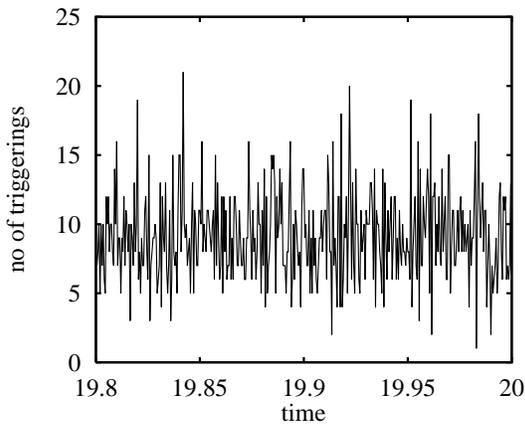}
 \caption{\label{Fig1} vo= 0.2,   width = 0.2,  k = 0.2:
  typical pattern of the triggering of the neurons in the semiclassical
  neural network. There is apparently no phase locking.}
\end{figure}

\begin{figure}[ht!]
\includegraphics[width=8cm]{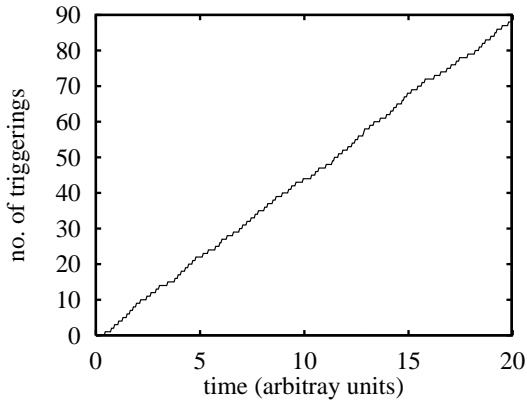}
 \caption{\label{Fig2}Cumulative number of triggering against time. One can
  see a fairly regular linear behavior despite quantum stochasticity.
  This is for a single chosen neuron.}
\end{figure}

\begin{figure}[ht!]
\includegraphics[width=8cm]{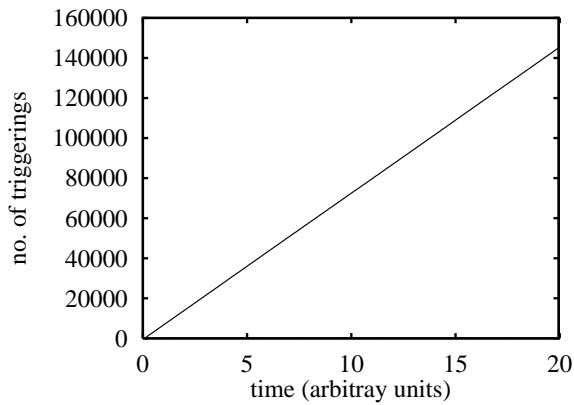}
 \caption{\label{Fig3}
 Same as Fig 2, but for the whole system.}
\end{figure}

Our simulations show that:

(1)  for a 40X40 lattice (as taken previously [1] in the classical
case) with periodic boundary conditions - to make it look like an
infinite lattice - for about 40,000 simulations for a given set of
parameters but different initial states, we get a constant average
for the neurons in the lattice. So in an average sense we do have
a period (Fig. \ref{Fig1}). The system is not chaotic. For a
single neuron, if we follow its history we see a fairly linear
relation between the cumulative number of firings and time (Fig.
\ref{Fig2}). If we consider the pooled sum of all nodes of the
lattice, the stochasticity almost disappears and we see a
practically straight line (Fig. \ref{Fig3}).


(2) Indeed, as the strength of the signal from the neighbor
increases, the time period    decreases (Table \ref{tab1}), but it
does not appear to go to zero. Unlike the classical case, here we
can make the signals arbitrarily strong without worrying about
running into a singularity or negative periods. The parameter $q$
was fitted to the first value and then Eq. \ref{eq8} seems capable
of predicting all the other periods with remarkable accuracy.

\begin{table}
\caption{\label{tab1}Strength of Quantum Potential and Average
Period of Neurons \\
\\$k=1$, $q= 1.4$, $width= 0.2$, $V_0= 1$}
\begin{ruledtabular}
\begin{tabular}{lcr}
$v$  &  $\overline{T_{pred}}$ & $\overline{T_{sim}}$ \\
 \hline
0.1 &   0.050   &  0.050 \\
0.2 &   0.035   &  0.035 \\
0.3 &   0.027   &  0.028 \\
0.4 &   0.023   &  0.023 \\
0.5 &   0.020   &  0.020 \\
1.0 &   0.013   &  0.013 \\
10.0 &  0.0028  &  0.028 \\

\end{tabular}
\end{ruledtabular}
\end{table}

(3) The width of the pulse is important here too. If the pulse is
spread out, the average period increases (Table \ref{tab2}). Again
we get excellent agreement of Eq. \ref{eq8} with the simulation
results for the periods, with the same value of $q$ used in Table
\ref{tab1}.
\begin{table}
\caption{\label{tab2}Variation of  Period with Duration of Quantum
Potential
\\$k=1$, $q=1.4$, $v= 0.2$, $V_0= 1$}
\begin{ruledtabular}
\begin{tabular}{lcr}
$width$  & $\overline{T_{pred}}$   &  $\overline{T_{sim}}$ \\
 \hline
0.1 &  0.023    &       0.023 \\
0.2 &  0.035    &       0.035 \\
0.3 &  0.044    &       0.043 \\
0.5 &  0.056    &       0.054 \\
1.0 &  0.074    &       0.065 \\

\end{tabular}
\end{ruledtabular}
\end{table}

\begin{figure}[ht!]
\includegraphics[width=8cm]{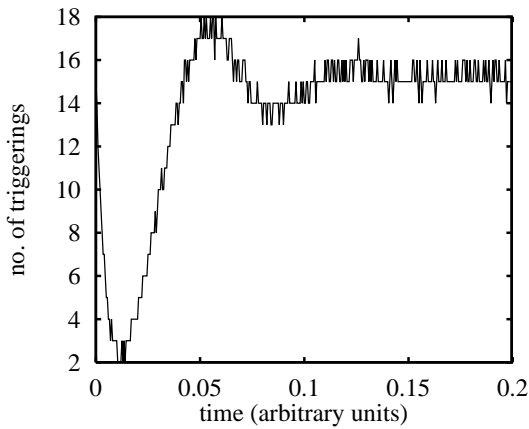}
 \caption{\label{Fig4}Transition from short term behavior to asymptotic
 behavior with all peripheral nodes initially in state $1$.}
\end{figure}

(4) The input dependence is observed by averaging over 100
simulation runs.  The following four types of inputs were used:

(a) All peripheral nodes in state $|1>$; all body nodes in state
$|0>$. In this case we see (Fig. \ref{Fig4}) a smooth transition
from a state with an initial firing rate proportional to the
number of initially excited nodes dying down quickly as the system
forgets the input and lets the system parameters take over with a
noisy pattern, despite the averaging over the runs. It is
remarkable that the initial few cycles with the memory show
virtually no noise.

\begin{figure}[ht!]
\includegraphics[width=8cm]{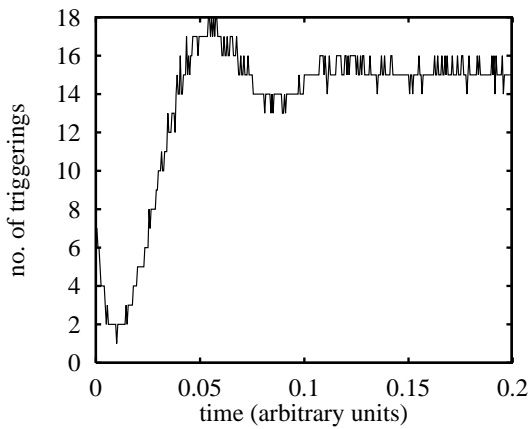}
 \caption{\label{Fig5}Same for peripheral nodes initially in states
  $|1>$ and $|0>$ alternately.}
\end{figure}

(b) Peripheral nodes alternately in states $|1>$ and $|0>$; body
nodes in state $|0>$ (Fig. \ref{Fig5}). In this case we begin with
a smaller number of firings on account of halving the initially
excited nodes, and there are a few kinks in the initial cycles,
possibly due to the conversion of the spatial lack of symmetry to
temporal. As expected the system moves to the common noisy
asymptotic behavior after forgetting the input.

\begin{figure}[ht!]
\includegraphics[width=8cm]{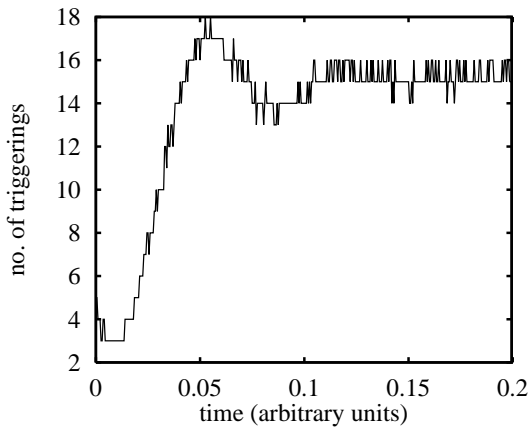}
 \caption{\label{Fig6}Same for peripheral nodes initially in random
 states of excitation.}
\end{figure}

(c) Peripheral nodes in random states between $|0>$ and $|1>$.
Here too we see fairly prominent kinks (slightly less than the
oscillating $|1> \Leftrightarrow |0>$ pattern in the previous
case) from the interaction of the noncoherent randomness of input
in the neighboring peripheral nodes until the short-term memory
disappears (Fig. \ref{Fig6}).

(d) In this case we take the whole lattice to be initially in state
$|0>$. However because of the external driving potential the system
soon develops into a noisy final state from a smooth initial state
with no excitation.

\section{Discussion}

We have seen that this semiclassical neural network which is
designed to mimic in some ways the neural network with an
integrate-and-fire model with classical action potentials
indicates the presence of an average time period reminiscent of
the classical model. The following differences stand out: 1) We
cannot have a system with a zero time period (infinitely fast). 2)
We can have arbitrarily high potentials linking the neurons.

We have seen that the average system period depends in a simple
way on the system parameters as given in Eq. \ref{eq8}. A single
parameter choice predicts a wide range of  potential pulse
strengths and durations.

An interesting possible extension of the ideas presented here
would be to use multi-state neurons (multibits)  with no classical
analogue and of course also to use coherent complex quantum action
to investigate the possibility of phase-locking from coherence.

We have not considered Hebbian learning by introducing
dissipation. Altaisky \cite{AL1} and Zak \cite{ZK1} have
investigated the effects of dissipation in quantum neural
networks. In our model also dissipation is inevitable as the
neurons are all allowed to fire independently, with no coherence
or entanglement. However, unlike these authors, we have considered
large systems where the role of adaptivity in the quantum context
becomes too complex for a preliminary investigation. The system,
nevertheless,  can serve as short-term memory with a built-in
mechanism for effacing all input dependence in the long run, which
may be a desirable characteristic where learning and unlearning
must go together. As the system we have considered is very basic,
it is possible that more complex systems may be designed for
greater functionality.

\begin{acknowledgements}
I would like to thank Andrew Tan at UCSF for useful discussions.
\end{acknowledgements}

\end{document}